\documentclass{iopart}
\usepackage{graphicx}

\usepackage[normalem]{ulem} 
\usepackage{color} 
\usepackage{subfig}
\usepackage{keyval}

\newcommand{\ket}[1]{|#1 \rangle}
\newcommand{\bra}[1]{\langle #1|}

\begin{document}
\title{Strategies for triple-donor devices fabricated by ion implantation.}

\author{Jessica~A.~Van~Donkelaar, Andrew~D.~Greentree,
Lloyd~C.~L.~Hollenberg and David~N.~Jamieson}

\address{Centre for Quantum Computer Technology, School of Physics,
The University of Melbourne, Melbourne, Victoria 3010, Australia.}




\begin{abstract}
Triple donor devices have the potential to exhibit adiabatic tunneling via the CTAP
(Coherent Tunneling Adiabatic Passage) protocol which is a candidate transport mechanism for
scalable quantum computing.  We examine theoretically the statistics
of dopant placement using counted ion implantation by employing an analytical 
treatment of CTAP transport properties under hydrogenic assumptions. 
We determine theoretical device yields for proof of concept devices 
for different implant energies.  In particular, we determine a significant 
theoretical device yield ($\sim 80\%$) for 14\,keV phosphorus in silicon with nominal $20~\mathrm{nm}$ spacing.
\end{abstract}




\section{Introduction}
Quantum computer (QC) architectures compatible with conventional silicon 
processing technologies would seem to be at an advantage over 
other schemes. This advantage is due to the perceived ability to leverage 
conventional silicon processing techniques. Of particular interest 
is the Kane solid-state quantum 
computer \cite{bib:KaneNature1998} which is based around nuclear 
spins of phosphorus ($^{31}$P) donors in an isotopically pure 
$^{28}$Si matrix (Si:P).  As all of the controls
and couplings are essentially electron spin mediated or effected,
an electron spin version version was proposed by Hill \emph{et al.}
\cite{bib:HillPRB2005} that took advantage of global control
properties to access the faster electron, rather than nuclear, spin 
gate times.

The Si:P QC still faces substantial technical challenges before it
can become a reality, and effective means of transport such as that
proposed by CTAP (Coherent Tunneling Adiabatic Passage)
\cite{bib:GreentreePRB2004} offer significant flexibility in
overcoming these challenges.  Amongst these challenges are the
expected penalty for linear nearest neighbour (LNN) architectures in
decreased threshold for fault-tolerant error correction due to extra
SWAP operations \cite{bib:Stephens,bib:Szkopek}, and with Si:P, valley
degeneracy leads to sensitivity of most critical parameters to
variations in donor positions at the atomic level
\cite{bib:KoilerPRL2002,bib:WellardPRB2005}.  Both these problems
can be ameliorated by the use of effective long-range transport
which allows the LNN bottleneck to be avoided, and the incorporation
of defect-tolerant design methodologies.  An architecture based
around transport via CTAP was recently proposed
\cite{bib:HollenbergPRB2006} and a threshold analysis performed
for a bilinear geometry \cite{bib:Stephens}.  The use of CTAP
transport rails as designed in \cite{bib:HollenbergPRB2006} also
avoids a complex \emph{classical} control issue, namely the
prohibitively high gate density of the original Kane scheme
\cite{bib:Oskin2003}.  Although the proof-of-concept structures that
we discuss here do not provide the advantages of CTAP rails,
demonstrating three-dopant devices is a critical step
towards eventual scalable structures.  The preferred method 
for fabricating longer CTAP chains is hydrogen resist lithography 
that allows for near-atomic dopant placement \cite{bib:schofield03}. This bottom-up process has recently had remarkable success in creating nanoscale surface structures \cite{bib:Rueb07,bib:small} for quantum devices. However the demonstration of the CTAP mechanism requires the construction of single dopant arrays deep within a silicon substrate. Currently, the quickest route to the construction of a proof-of-principle device is the top-down method of single ion implantation \cite{bib:JamiesonAPL2005} as it can be adapted to such fabrication with existing technologies. Here we investigate this practical method for engineering a 
three atom device. We show estimates for the times required for high-fidelity CTAP
using hydrogenic approximations to the tunnel matrix elements \cite{bib:OpenovPRB2004} 
and ion implanted donor positions calculated using the SRIM
\cite{bib:SRIM} package. Analytic solutions to the expected time for
CTAP based on the adiabaticity criterion allow some statistics of
expected device yields to be made, which are of benefit in guiding
experimental investigations.  We also discuss some of the
limitations of our approximations, and point to some of the
challenges in identifying complete adiabatic pathways to realise the
CTAP protocol.

\section{Coherent Tunneling Adiabatic Passage and Adiabaticity}

CTAP is a protocol for the spatial transport of a particle between
two-points on a quantum chain.  In its simplest case this is a
three-site protocol with the central chain being a single site. CTAP
is the direct spatial analog of the well-known STIRAP protocol from
quantum optics \cite{bib:Vitanov2001}.  CTAP is distinguished from
STIRAP in that with CTAP, all variations in couplings are effected
by direct modulation of the wave-function overlaps by either surface
gate control or well-proximity. With STIRAP the tunnel
matrix elements are strictly electromagnetic field driven
\cite{bib:STIRAPDD}. In addition to the electronic transfer
exploited for Si:P systems, CTAP has been proposed in atomic
lattices \cite{bib:EckertPRA2004}, Cooper-Pair boxes
\cite{bib:SiewertAdvSSP2004}, quantum-dots
\cite{bib:GreentreePRB2004,bib:PetrosyanOptComm2006}, Bose-Einstein
condensates \cite{bib:GraefePRA2006,bib:Rab2007}, spin chains
\cite{bib:Ohsima2007} and for photonic transport through coupled
waveguides \cite{bib:LonghiJPB2007}.  The latter case has also been
demonstrated in beautiful experiments by Longhi and co-workers
\cite{bib:Longhi2007,bib:Della2008} which adds significant impetus to the push to
demonstrate CTAP in systems of massive particles.

CTAP is also an example of a growing search for direct analogues of
existing quantum optical effects in quantum electronic systems, and
as the degree of coherence that can be detected increases one should
expect more sophisticated effects to be observed. Other than CTAP,
there are also proposals for Autler-Townes measurements
\cite{bib:GreentreePRB2004AT,bib:GreentreePRB2005}, and
coherent-population trapping-like dark state transport protocols
\cite{bib:MichaelisEPL2006,bib:Emary}. One attraction of quantum
electronics over conventional quantum optics lies  in the ability to
tailor the Hilbert space by construction, rather than being limited
by the structures of atomic systems.  This flexibility allows new
vistas and extensions to be explored, for example the
multiple-recipient adiabatic passage of
Refs.~\cite{bib:GreentreePRA2006,bib:DevittQIP2007} and the direct
spatial analogue of the tripod atom \cite{bib:Deasy}.

\begin{figure}[tb!]
\subfloat{\label{fig:1a}
\includegraphics[width=0.5\linewidth]{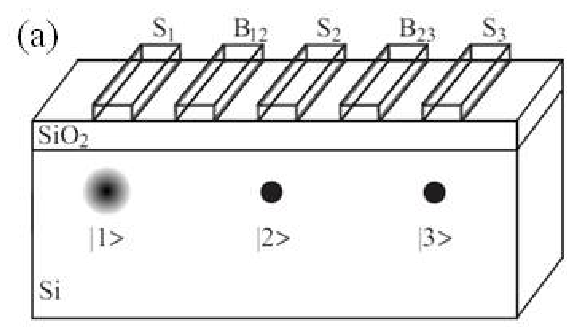}}\\
\subfloat{\label{fig:1b}
\includegraphics[width=0.5\linewidth]{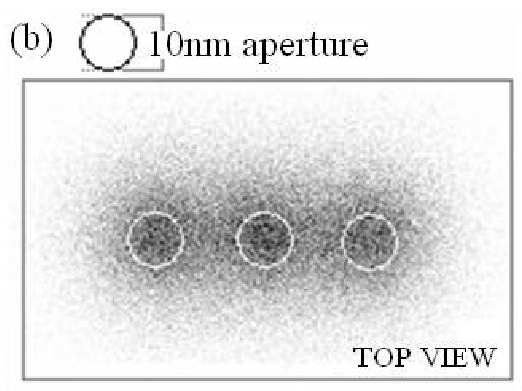}}\\
\subfloat{\label{fig:1c}
\includegraphics[width=0.5\linewidth]{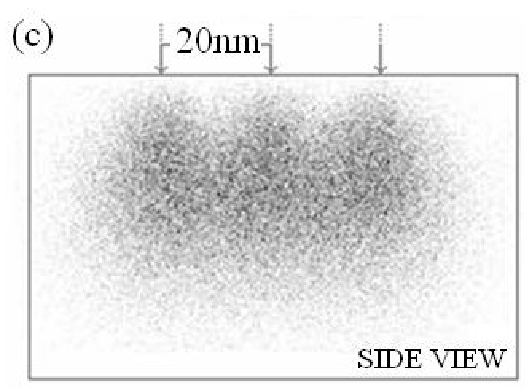}}\\
\caption{\label{fig:Fig1} (a) Triple dopant, one charge system with
surface gate control for realising CTAP. The symmetry gates $S_1$
and $S_3$ maintain the degeneracy of the end of chain states, whilst
the variations in the tunnel matrix elements is effected by the
barrier gates $B_{12}$ and $B_{23}$. (b) \textit{top veiw} 
(c) \textit{side view} SRIM simulations showing
the spatial probability distribution of 100,000 14\,keV phosphorus ions implanted into
silicon through three apertures 10~nm in diameter spaced 20\,nm apart.}
\end{figure}

To effect the CTAP pathway requires the adiabatic transformation of
one of the eigenstates of the system from a known start state to a
desired final state. In the spatial setting we restrict the control
parameters to the tunnel matrix elements which are controlled by
surface electrodes (gates), analogously to the way in which
electro-magnetic field intensities are varied in STIRAP. 
A schematic of the surface structure that we are assuming is shown in
Figure~\ref{fig:Fig1}\subref{fig:1a}, viewed from the top down. The device structure
shown in Figure~\ref{fig:Fig1}\subref{fig:1a} is yet to be made and it is the purpose
of this paper to investigate potential fabrication methods based on 
presently available technologies. Devices of comparable complexity 
have been fabricated using the counted-ion implantation techniques described in
Ref.~\cite{bib:JamiesonAPL2005}, and we also note the demonstration
of triple dots in GaAs 2DEG structures \cite{bib:GaudreauPRL2006,bib:SchroerPRB2007,bib:rogge2008} 
and gated carbon nanotubes \cite{bib:grove2008}.\\ 

Ideally we would like three donor atoms spaced 20-30\,nm apart, 20\,nm below the surface. The SRIM data in Figures \ref{fig:Fig1}\subref{fig:1b} and \ref{fig:Fig1}\subref{fig:1c} shows the probability distribution of 100,000 14\,keV P$^+$ ions implanted sucessively through three circular apertures 10\,nm in diameter. A 14\,keV phosphorous ion will travel around 20\,nm into the substrate before stopping but can straggle away from this median position by up to 11\,nm as it undergoes collisions with the Si lattice. A less energetic 7\,keV ion will penetrate 14\,nm below the surface and has 40\% less straggle than at 14\,keV. The straggle imposes constraints on the accuracy of dopant placement. To quantify the consequence of these fabrication perturbations on the ideal donor positions seen in Figure \ref{fig:Fig1}\subref{fig:1a} we have used SRIM simulations with a simplified treatment of three-donor CTAP.\\

We begin by writing down the Hamiltonian of the three
donor one electron problem in the three-state approximation (i.e.
where we only keep the lowest state of the electron localised around
the donor).  In the basis $\ket{1}, \ket{2}, \ket{3}$ with onsite
energies $E_i = 0$ (assuming fully compensated) and tunnel matrix
elements $\Omega_{12}(t)$ and $\Omega_{23}(t)$, we also include the
possibility of next nearest neighbour tunneling (i.e. from $\ket{1}$
to $\ket{3}$) however for the discussion that follows this will be
assumed to be zero.  The Hamiltonian is
\begin{eqnarray}
\mathcal{H} = \Omega_{12}(t)\ket{2}\bra{1}
            + \Omega_{23}(t)\ket{3}\bra{2}
            + \Omega_{13}(t)\ket{3}\bra{1} + h.c.,
            \label{eq:CTAPHam}
\end{eqnarray}
and with $\Omega_{13}(t) = 0$, the eigenvalues are
\begin{eqnarray}
\ket{\mathcal{D}_0} &=& \frac{\Omega_{23}\ket{1} -
\Omega_{12}\ket{3}}{\sqrt{\Omega_{12}^2 +
\Omega_{23}^2}}, \\
\ket{\mathcal{D}_{\pm}} &=&
 \frac{\Omega_{12} \ket{1} \pm
\sqrt{\Omega_{12}^2 + \Omega_{23}^2} \ket{2} + \Omega_{23}
\ket{3}}{\sqrt{2\left(\Omega_{12}^2 + \Omega_{23}^2\right)}} ,
\end{eqnarray}
with energies
\begin{eqnarray}
E_0 &=& 0, \\
E_{\pm} &=& \pm \sqrt{\Omega_{12}^2 + \Omega_{23}^2}.
\end{eqnarray}
The CTAP protocol can now be understood quite simply. The idea is to
remain in the state $\ket{\mathcal{D}_0}$, and to vary the tunnel
matrix elements so that at time $t=0$, the system is in the desired
initial state, e.g. $\ket{\mathcal{D}_0 (t = 0)} = \ket{1}$, and at
time $t=t_{\max}$ the system is in the desired final state, e.g.
$\ket{\mathcal{D}_0 \left(t = t_{\max}\right)} = \ket{3}$. 
Note that although this three mode description for the tunneling is 
obviously a simplification, it still captures all of the essential 
physics of the CTAP protocol, a fact confirmed by recent analyses of 
the CTAP in the triple square well case \cite{bib:cole2008}.

\begin{figure}[tb!]
\includegraphics[width=0.5\linewidth]{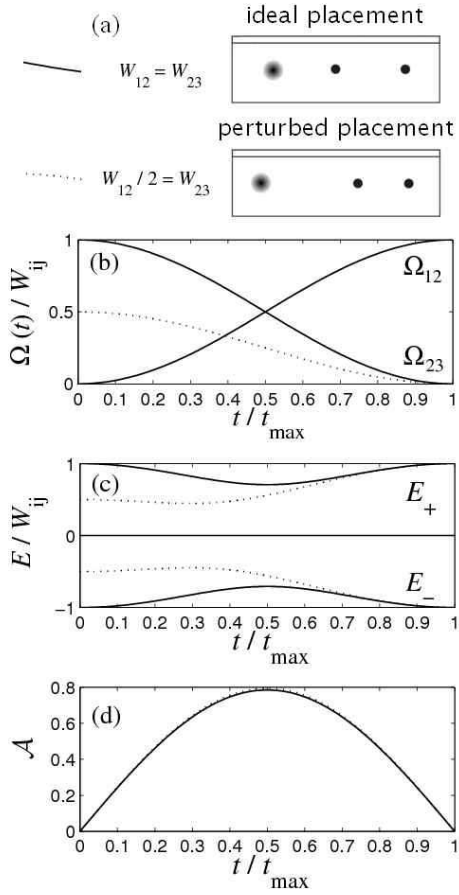}
\caption{\label{fig:CTAPfig} (a) Two possible scenarios for implanted triples. (b) Tunnel matrix elements as a
function of time.  The solid lines correspond to the case that
$W_{12} = W_{23}$ for the sinusoidal variation as defined in the
text, whereas the dotted line corresponds to the case that $W_{23} =
W_{12}/2$. (c) Eigenenergies of the states as a function of time
with the sinusoidal variation, again solid lines are for $W_{12} =
W_{23}$ and the dashed to $W_{23} = W_{12}/2$. (d) the value of the
adiabaticity parameter $\mathcal{A}$ throughout the process. Note
that it is maximised at $t=t_{\max}/2$ irrespective of the values of
$W_{12}$ and $W_{23}$, although there are minor differences in the
value of $\mathcal{A}$.  As expected, the process is slightly less
adiabatic with the smaller values of $W$. }
\end{figure}

To effect CTAP, there is clearly a large amount of flexibility to
choose the pulsing scheme for the tunnel matrix elements.  In STIRAP
protocols, gaussian or gaussian-like pulses are most commonly
employed because of the necessity to turn on the excitation (laser
pulse) before varying it \cite{bib:Vitanov2001}. However in the
solid state, where non-zero tunnel matrix elements arise solely due
to donor proximity, this necessity is not required, and so we
advocate the use of the pulses that vary between their extrema at
$t=0$ and $t = t_{\max}$.  Such pulses are explicitly stated below
and illustrated in Figure~\ref{fig:CTAPfig}. In
Refs.~\cite{bib:Rab2007,bib:DevittQIP2007,bib:GreentreeSPIE2005}
error function pulses were employed which have some advantages in
terms of smoothness of evolution and in nonlinear systems avoid
certain complications due to eigenstate degeneracy at the ends of
the protocol.  For simplicity, here we choose sinusoids with the
sole purpose of making the analytical results clearer.

To determine whether the system remains in the target state, we
invoke the adiabaticity criterion, and due to symmetry we define
(without loss of generality) the adiabaticity parameter to be between
$\ket{\mathcal{D}_0}$ and $\ket{\mathcal{D}_+}$.  The adiabaticity
parameter is
\begin{eqnarray}
\mathcal{A} = \frac{\bra{\mathcal{D}_+}\frac{\partial
\mathcal{H}}{\partial t}\ket{\mathcal{D}_0}}{\left| E_+ - E_0\right|
^2},
\end{eqnarray}
and for adiabatic evolution we require $\mathcal{A} \ll 1$. One
should be mindful of the fact that the adiabaticity does not
translate to a direct measure of fidelity, it is rather a measure
for when the assumption of adiabatic evolution is justified.
Choosing as the form for the (gate controlled) tunnel matrix
elements
\begin{eqnarray}
\Omega_{12}(t) &=&
 W_{12} \sin^2\left(\frac{\pi t}{2 t_{\max}}\right), \\
\Omega_{23}(t) &=&
 W_{23} \cos^2\left(\frac{\pi t}{2 t_{\max}}\right),
\end{eqnarray}
gives
\begin{eqnarray}
\mathcal{A} = \frac{\pi W_{12} W_{23} \sin \left(\frac{\pi
t}{t_{\max}}\right)}{\sqrt{2} t_{\max} \left(W_{12}^2 +
W_{23}^2\right)^{3/2}}. \label{eq:Acrit}
\end{eqnarray}
Of importance here is to note that the adiabaticity is a maximum
when $t=t_{\max}/2$ irrespective of the relative values of
$W_{12}$ and $W_{23}$. This is significant in the design of robust
sequences and in affording a quick estimate of the required
timescale for CTAP operation.  To illustrate the energies, tunnel
matrix elements and adiabacity, in Figure~\ref{fig:CTAPfig} we present
characteristic plots for the case that $W_{12} = W_{23}$ and when
$W_{23} = W_{12}/2$.

In a realistic experiment, we will want to set the time for CTAP
given a certain desired adiabaticity. So it is more important to
rearrange (\ref{eq:Acrit}) to determine the value for $t_{\max}$
that keeps the maximum value of $\mathcal{A}$ at or below some
threshold, which is
\begin{eqnarray}
t_{\max} = \frac{\pi W_{12} W_{23}}{\sqrt{2} \mathcal{A}
\left(W_{12}^2 + W_{23}^2\right)^{3/2}}. \label{eq:Timecrit}
\end{eqnarray}
where we have dropped the sine term because we are evaluating
$t_{\max}$ with respect to the maximum value of $\mathcal{A}$.

Equation (\ref{eq:Timecrit}) is particularly useful in gaining insight into
the practicalities of CTAP.  Although we are not yet in a position
to place a lower bound on $t_{\max}$ (in general this must come from
a detailed understanding of the higher lying molecular and orbital
states,  limitations on the pulsing electronics, and accessible
surface gate voltage), we can immediately compare our results with
realistic upper bounds. The most obvious impediment to very long
time scales will be the limits placed by decoherence, which will set
a maximum length of time over which the protocol can be conducted
\cite{bib:GreentreePRB2004,bib:IvanovPRA2004,bib:twamley2008}.

Until now we have neglected the next nearest neighbour tunneling,
i.e. $\Omega_{13}$.  Formally there is no CTAP pathway for non-zero
$\Omega_{13}$ as the Hamiltonian (\ref{eq:CTAPHam}) has no null
(or dark) state, however we can place a good bound on whether
neglecting $\Omega_{13}$ will be valid by comparing the period for
oscillation on the $\ket{1}-\ket{3}$ transition directly with the
total time for CTAP.  We introduce $\mathcal{J} \equiv \Omega_{13}
t_{\max}$ and assert that if $\mathcal{J} \ll 1$ then we may ignore
the effect of $\Omega_{13}$ in our analysis. This criterion is
helpful for proof of concept devices, but may not suffice for full
QC applications where more rigorous error control is required
\cite{bib:HollenbergPRB2006}.

\section{Implications for ion implanted devices}

We have treated the Hamiltonian in an ideal case, without regard to
the physical nature of our system.  If we now turn specifically to phosphorous
dopants in silicon, with realistic gate controls, then issues of the
suppression of tunnel matrix elements via barrier control,
cross-talk between gates \cite{bib:KandasamyNano2006} and
microscopic details of the dopant species become paramount
\cite{bib:KoilerPRL2002}. All such features must be taken into
consideration to determine the adiabatic pathway, which is a highly
non-trivial task which we do not attempt here.  Instead we have
focussed on hydrogenic approximations which allow us to gain rapid
insight into a variety of implant conditions; we leave
the solution of the control problem to future work, assuming (as in
Refs.~\cite{bib:GreentreePRB2004,bib:HollenbergPRB2006}) that the
gate controls are able to completely suppress the tunnel matrix
element from the maximum (no-field) level to zero.

To determine the bare (unperturbed) tunnel matrix elements, we use
the hydrogenic approximation of Openov \cite{bib:OpenovPRB2004}, 
an approach used ostensibly in the study of singly ionized double-donor 
structures \cite{bib:tsukanov07}.  These results are applicable to donor 
separations predominantly along the [100] direction in silicon 
where the valley degeneracy has less effect \cite{bib:KoilerPRL2002,bib:WellardPRB2006}. 
To be explicit, we use

\begin{eqnarray}
W_{ij} = 4 E^*\left(\frac{d_{ij}}{a^*_B}\right)
\exp\left(-\frac{d_{ij}}{a^*_B}-1\right),
\end{eqnarray}
$E^*$ is the effective Hartree, and $d_{ij}$ is the interdonor
separation between donor $i$ and $j$.  With these constraints we can
immediately write down $t_{\max}$ and $\mathcal{J}$ as
\begin{eqnarray}
 t_{\max}&=& \frac{\pi a^*_B d_{12}d_{23}
 \exp\left(-\frac{d_{12}+d_{23}}{a_B^*}-2\right)}
 {4 \sqrt{2} E^* \mathcal{A}
 \left\{ \left(F_{12}\right)^2 + \left(F_{23}\right)^2 \right\}^{3/2}}, \\
 \mathcal{J}&=& \frac{\pi d_{12}d_{23} d_{13} \exp\left(-
\frac{d_{12}+d_{23}+d_{13}}{a_B^*} -3\right)}{\sqrt{2} \mathcal{A}
\left\{ \left(F_{12}\right)^2 + \left(F_{23}\right)^2 \right\}^{3/2}}, \\
F_{ij} &=& d_{ij} \exp\left( - \frac{d_{ij}}{a^*_B} - 1\right)
 = \frac{W_{ij} a_B^*}{4 E^*}. 
\end{eqnarray}
A rigorous calculation of $t_{\max}$ and $\mathcal{J}$ would involve
full band structure considerations, but these formulae give an
extremely efficient mechanism for determining these important
parameters with relatively minimal computational cost, as is
required to process the large number of dopant positions that can be
obtained using SRIM calculations.

To explore the range of expected yields, we performed SRIM simulations for 
14\,keV and 7\,keV phosphorus. Single 14\,keV phosphorous ions are 
routinely detected entering a silicon substrate \cite{bib:JamiesonAPL2005} 
and reach a nominal depth of 20\,nm below the substrate surface. The lateral 
straggle of an implanted ion limits the precision of dopant placement 
and hence the control of donor seperation required to engineer 
a suitable proof of concept device. Therefore 7\,keV was also 
chosen for the greater placement accuracy it allows and as a guide 
to future developement of the detection system. A typical oxide thickness 
of 5\,nm was used in the 14\,keV strategy while in the low energy 
strategy we assumed an oxide thickness of an optimal 1.2\,nm.
The on-chip detection of \cite{bib:JamiesonAPL2005} guarantees that
precisely three ions will be implanted, however it does not ensure
that there will be one ion per aperture. To ensure this capability a 
{`}step-and-repeat{'} strategy based on existing technologies is currently 
being integrated with the detection system. A similar approach is being 
followed by Schenkel et al. \cite{bib:persaud05,bib:schenkel06} and also by Meijer et al. \cite{bib:Meijer08} for the creation of nitrogen-vacancy colour centres in diamond. 

\begin{figure}[tb!]
\includegraphics[width=0.5\columnwidth]{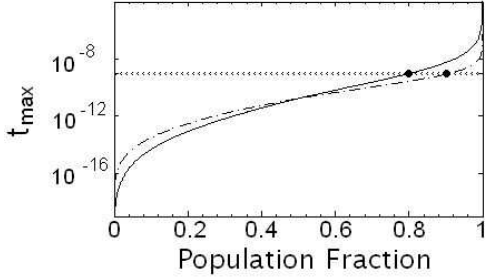}
\caption{\label{fig:CTAPMetrics} Cumulative distribution functions
showing the fraction of implant triples with CTAP times at
adiabaticity $\mathcal{A} = 0.01$ for different implant strategies.
The solid line is for 14~keV P implanted through 5~nm of SiO$_2$ on
Si; the dashed line is for 7~keV P through
1.2~nm of SiO$_2$ on Si, a practical
lower limit to conventional fabrication. The horizontal line
indicates $10^{-9}$ seconds, a realistic limit on the CTAP transfer time due to dephasing. From this we may
infer that the 14\,keV implant strategy has an expected yield around $80\%$,
whilst the 7\,keV strategy has an expected yield around $90\%$. Note that
these yields do not take into account the \emph{lower} bounds on
controllability, and some fraction of dopant triples will have
tunnel matrix elements that are too large to be effectively controlled.}
\end{figure}

The results of performing SRIM simulations for each implant strategy
are shown in Figure~\ref{fig:CTAPMetrics}.  In addition to the raw
straggle data obtained from SRIM, we have convolved the
distributions with apertures 10~nm in diameter, separated linearly
by $20~{\rm nm}$.  For each implant triple, we have then calculated
$t_{\max}$ and $\mathcal{J}$, although as $\mathcal{J}$ is found to
be substantially less than one for over $95\%$ of
the triples we have not shown those results here.
Figure~\ref{fig:CTAPMetrics} then shows the cumulative distribution
function for the $t_{\max}$ at an adiabaticity of $\mathcal{A} =
0.01$, which may immediately be interpreted as a measure of yield.
Upper limits on $t_{\max}$ are given by the decoherence limit.  As a
rule of thumb, we wish to be at least an order of magnitude faster
than decoherence for a significant proof-of-concept signal
\cite{bib:GreentreePRB2004}.  To aid comparisons,
Figure~\ref{fig:CTAPMetrics} shows a horizontal line corresponding to
1\,ns, which is an order of magnitude less than a realistic expected decoherence time (10\,ns). 
From this we can read off the population of triples which satisfy this criterion for
each implant strategy.  One should be mindful, however, that the
expected device yield \emph{even without fabrication errors} will be
less than this, due to the fraction of donors that are too close
(i.e. have too high tunnel matrix elements) to be properly
controlled, or to have their tunnel matrix elements adequately
suppressed.  This bound will be determined in large part by the
breakdown of the oxide barrier and is not addressed here.

From the results in Figure~\ref{fig:CTAPMetrics} we can make some
initial estimates of the prospects for ion implantation strategies,
and in particular we find that within the limits of our models, we
can afford some optimism about the ion implantation as a strategy
for proving the concept of CTAP.  The expected theoretical yield for the standard 14\,keV
strategy is predicted to be around $80\%$, and for the low energy 7\,keV strategy
it is expected to be closer to $90\%$.  

\section{Conclusions}

We have performed analysis of implanted donor triples using SRIM to
provide guidance to experimental efforts to realise the CTAP
(Coherent Tunneling Adiabatic Passage) protocol.  Our results
suggest theoretical device yields around $80\%$ within the
hydrogenic approximation and not including device fabrication
errors.  To process the large amounts of data generated by SRIM, we
developed new analytical results for the required time for
high-fidelity CTAP, and applied these metrics to various
implantation strategies and species.  Our methods are unable to
address all of the requirements for CTAP, which include the need to
identify the particular adiabatic pathway for a given microscopic
location of donors, nor do we go beyond the hydrogenic
approximation.  More accurate modelling is required, (e.g. NEMO3D
\cite{bib:NEMO,bib:NEMOCTAP}) to theoretically describe a given
system, but this is far too numerically intensive to explore the
very large number of configurations explored here.  On the basis of
our results, we predict that ion implantation is a sensible strategy
to exploit for proof-of-concept CTAP devices, although the overall
scaling to the longer CTAP chains required for scalable QC
\cite{bib:HollenbergPRB2006} is expected to be far less favorable 
and hydrogen resist lithography methods for producing such CTAP 
chains is the preferred method \cite{bib:schofield03}.

\ack
The authors thank L. Jong for useful discussions. ADG is the recipient of an Australian Research Council Queen Elizabeth 
II Fellowship (project numberDP0880466), and LCLH is the recipient of an 
Australian Research Council Australian Professorial Fellowship 
(project number DP0770715). We acknowledge financial support from the Australian Defence Science and
Technology Organisation, the Australian Research 
Council, the Australian Government and US Army Research Office under Contract No. W911NF-04-1-0290.\\

\begin{center}

\end{center}

\end{document}